\def\year{2021}\relax
\documentclass[letterpaper]{article} 
\usepackage{aaai21}  
\usepackage{times}  
\usepackage{helvet} 
\usepackage{courier}  
\usepackage[hyphens]{url}  
\usepackage{graphicx} 
\usepackage{graphicx}  
\usepackage{natbib}  
\usepackage{caption} 
\usepackage{balance}

\usepackage{wrapfig}

\usepackage{caption}
\usepackage[separate-uncertainty=true]{siunitx}
\usepackage{mathtools}
\usepackage{graphicx}
\usepackage{courier}
\usepackage{epstopdf}
\usepackage{color}
\usepackage{csquotes}

\usepackage{amsmath}
\usepackage{amsfonts}
\usepackage{amssymb}
\usepackage[subnum]{cases}
\usepackage{booktabs} 
\usepackage{pifont}

\usepackage{algorithm,algorithmicx,algpseudocode}

\usepackage{amssymb}
\usepackage{pifont}

\usepackage {tikz}
\usetikzlibrary {positioning}
\definecolor {processblue}{cmyk}{0.96,0,0,0}
\usepackage{lipsum,adjustbox}

\usepackage[subnum]{cases}
\usepackage{color,soul}

\usepackage{subcaption}

\usepackage{footnote}
\makesavenoteenv{tabular} 

\usepackage{multirow}

\usepackage{booktabs,subcaption,amsfonts,dcolumn}
\newcolumntype{d}[1]{D..{#1}}

\usepackage{xcolor}

\newtheorem{lemma}{Lemma}

\usepackage{textcomp}

\title{Graph Neural Network to Dilute Outliers for Refactoring Monolith Application}
\author{

    Utkarsh Desai, Sambaran Bandyopadhyay, Srikanth Tamilselvam
}
\affiliations{

    IBM Research\\

    udesai26@in.ibm.com, samb.bandyo@gmail.com, srikanth.tamilselvam@in.ibm.com

}

\begin{document}

\maketitle

\begin{abstract}
Microservices are becoming the defacto design choice for software architecture. It involves partitioning the software components into finer modules such that the development can happen independently. It also provides natural benefits when deployed on the cloud since resources can be allocated dynamically to necessary components based on demand. Therefore, enterprises as part of their journey to cloud, are increasingly looking to refactor their monolith application into one or more candidate microservices; wherein each service contains a group of software entities (e.g., classes) that are responsible for a common functionality. Graphs are a natural choice to represent a software system. Each software entity can be represented as nodes and its dependencies with other entities as links. Therefore, this problem of refactoring can be viewed as a graph based clustering task. In this work, we propose a novel method to adapt the recent advancements in graph neural networks in the context of code to better understand the software and apply them in the clustering task. In that process, we also identify the outliers in the graph which can be directly mapped to top refactor candidates in the software. Our solution is able to improve state-of-the-art performance compared to works from both software engineering and existing graph representation based techniques.


\end{abstract}


\section{Introduction}\label{sec:intro}
Microservices is an architectural style that structures an application as a set of smaller services \cite{fowler}. These services are built around business functionalities and follow “Single Responsibility Principle”\footnote{https://www.infoq.com/articles/microservices-intro}. This allows the team to develop business functionalities independently. Also, they naturally benefit from cloud deployment due to the support for differential and dynamic addition of resources like CPU, memory, disk space to specific services based on the demand. However, there are lot of existing monolith applications currently in use that cannot fully tap these benefits due to their architecture style. Monoliths package all the business functionalities into a single deployable unit making them unsuitable to fully leverage cloud benefits. Therefore, there is a surge in enterprises wanting to refactor their monolith applications into microservices. This is done by mapping business functions onto the code structure and identifying the functional boundary in such a way that there are less dependencies across the services \cite{jin2019service}. In typical monoliths, there are classes (or programs) loaded with overlapping functionalities \cite{kalske2018transforming}. This can be identified by their dependencies with cross functional classes. We refer to such classes as outliers or refactorable candidates. They typically require top attention from the developers for modification during refactoring to make the microservices independent and deployable. 
But identifying functional boundaries on the existing code is a hard task \cite{gouigoux2017monolith} and the effort gets multiplied when done without the help of original developers, which is typically the case with legacy applications. 

\begin{figure}[!t]
  \centering
  \includegraphics[width=\linewidth]{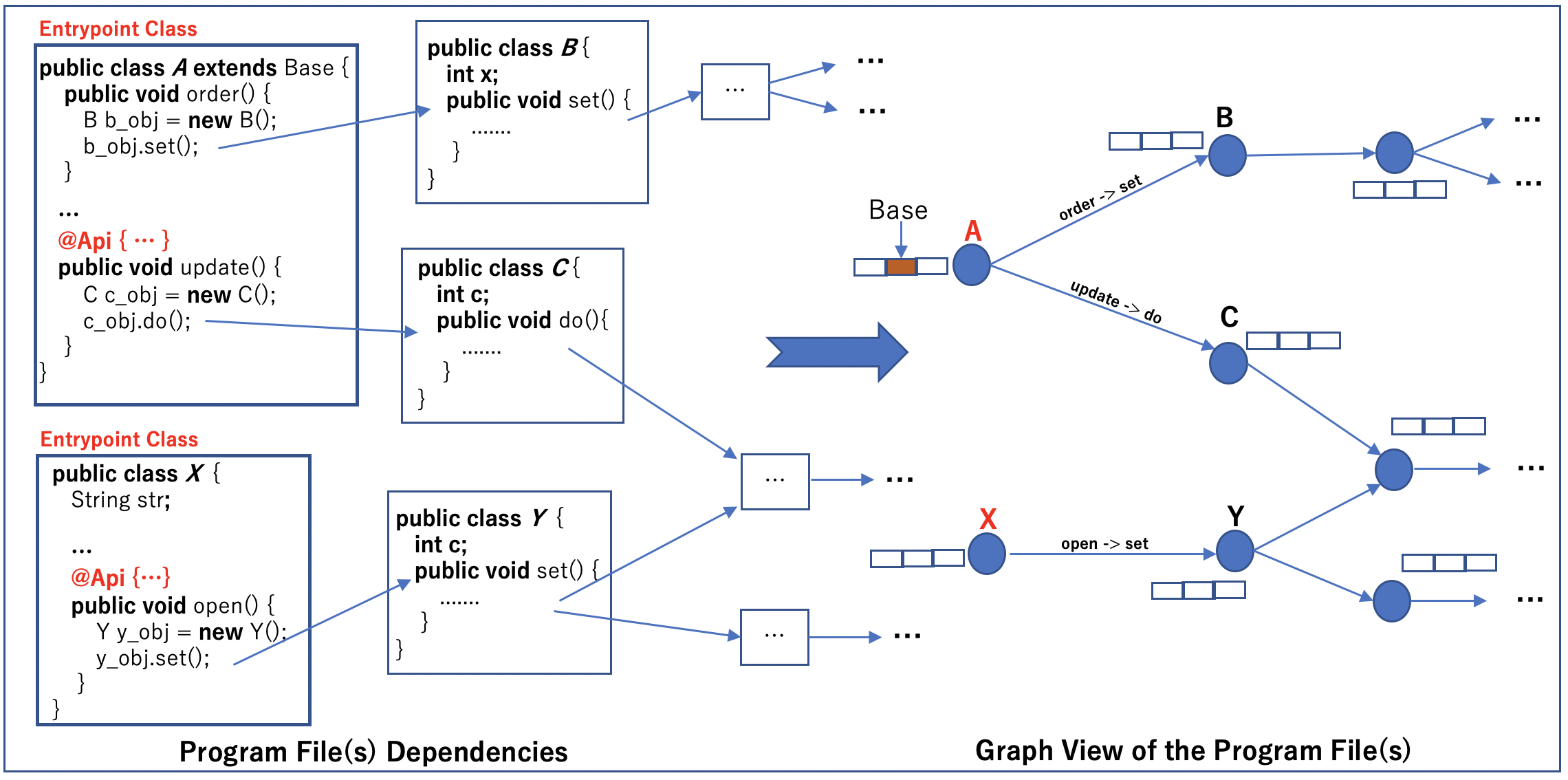}
  \caption{Representation of a sample Java application as graph: The method \textit{order()} from class \textit{A} invokes the method \textit{set()} from class \textit{B}, establishing a direct relation between the two classes. If we represent classes \textit{A} and \textit{B} as nodes in a graph, we can define a directed edge, $e(A,B)$ from \textit{A} to \textit{B}.}
  \label{fig:code_graph}
\end{figure}

In the software engineering community, the problem is often referred as software decomposition and several approaches~\cite{mono2micro-survey-guide} have been proposed. The approaches range from process mining, genetic algorithms to graph based clustering. Graphs are a natural way to represent application implementation structure. The classes in the application can be considered as nodes and its interaction with the other classes can be considered as edges. Further, the nodes can carry multiple features based on their type and their invocation pattern. Figure \ref{fig:code_graph} demonstrates the translation of an application into a graph. Therefore, the application refactoring problem can be viewed as a graph based clustering task. In the past, many clustering techniques have been applied on code \cite{clustering-in-se-survey}, but they often consider only the structural features of the application i.e the dependency of classes. Also, none of these approaches have looked into attributed graph networks or attempted to minimize the effect of outlier nodes during clustering.




Graph based mining tasks have received significant attention in recent years due to development of graph representation learning that maps the nodes of a graph to a vector space \cite{perozzi2014deepwalk,hamilton2017representation}. They have also been applied to a diverse set of applications such as social networks \cite{kipf2017semi}, drug discovery \cite{gilmer2017neural}, transportation and traffic networks \cite{guo2019attention}, etc.
In this work, we propose a novel graph neural network based solution to refactor monolith applications into a desired number of microservices. The main contributions of our paper are listed below.
\begin{enumerate}
    \item We propose a novel way to translate the application implementation structure to an attributed graph structure through static program analysis.
    \item We introduce two types of outliers that reflect the top refactoring program candidates.
    \item We propose a novel graph neural network (GNN), referred as CO-GCN\footnote{Code available at: https://github.com/utkd/cogcn} (\underline{C}lustering and \underline{O}utlier aware \underline{G}raph \underline{C}onvolution \underline{N}etwork), which unifies node representation, outlier node detection \& dilution and node clustering into the same framework for refactoring monolith applications.
    \item We improve the state-of-the-art performance with respect to both software engineering and graph representation based techniques to refactor four publicly available monolith applications. We attach the source code in the supplementary material for reproducibility of the results. 
    
\end{enumerate}



\section{Related Work}\label{sec:rel}
Fritzsc et al.~\cite{mono2micro-survey-guide} presented a survey
on ten different approaches towards refactoring a monolith application into microservices. Of these, only four works were applied directly on application code and the rest used other application artefacts such as logs, commit histories, UML diagrams etc. 
However, all of these works have drawbacks since they either (1) focus on only structural features; or (2) propose partitions focusing more on technical layers which is not desirable \cite{ms-badsmells}; or (3) partition only a subset of program files like EJBs in java.
\cite{mazlami-icws-2017} proposed a graph based clustering approach with a focus on version history. \cite{jin2019service} proposed hierarchical clustering of program files, but requires access to the runtime behavior of application which is practically difficult.
Moreover, these approaches do not exploit the power of representation learning and graph neural networks.
Also, they do not recommend refactorable classes.

Graph representation learning \cite{hamilton2017representation} shows promising results on multiple downstream graph mining tasks. Graph neural networks \cite{wu2020comprehensive} apply neural network directly on a graph structure. In Graph convolution networks introduced by \cite{kipf2017semi}, a localized first-order approximation of spectral graph convolutions is proposed and experimented for semi-supervised node classification. An unsupervised variation GCN autoencoder is proposed in \cite{kipf2016variational}. 
GNNs are also proposed for supervised \cite{chen2018supervised} and unsupervised community detection \cite{zhang2019attributed} in graphs. Recently, a self-supervised learning based GNN, Deep Graph Infomax (DGI) \cite{velivckovic2018deep} is proposed for obtaining node representation by using the principle of information maximization. Outlier nodes are present in any real-world graph. Outliers are shown to have adverse effect on the embeddings of regular nodes in a graph \cite{liang2018semi}. Unsupervised algorithms to minimize the effect of outliers in the framework of graph representation learning are proposed recently \cite{bandyopadhyay2019outlier,bandyopadhyay2020outlier,bandyopadhyay2020integrating}. However, minimizing the effect of outliers in the framework of GNN has not been addressed in the literature.

\section{Methodology}\label{sec:soln}
Given a monolith application, we want to partition the monolith into $K$ clusters of classes, with $K$ provided by a subject matter expert (SME), where each cluster is a group of classes that perform a well-defined functionality. The clusters should exhibit high cohesion, i.e., have strong interaction within the cluster and low coupling i.e., less interaction between clusters. We also want to identify the following outlier classes from a monolith application \cite{bandyopadhyay2019outlier} to be handled by an SME. 
\begin{itemize}
    \item \textit{Structural Outlier}: A class which has high interaction with classes from different clusters.
    \item \textit{Attribute Outlier}: A class which has attributes, such as usage patterns, similar to attributes from other clusters.
\end{itemize}

\subsection{Converting Applications to Graph}\label{sec:convGraph}
We now describe our approach to represent an application as a graph, given it's source code. Consider a simple Java application comprising of multiple classes as shown in Figure \ref{fig:code_graph}. Each class in the application can be represented as a node in a graph. We denote the set of such nodes as $V$. We establish a \textit{directed} edge from node A to node B if there is method in the class A that calls a method from class B. We perform static analysis\footnote{https://github.com/soot-oss/soot} of the application code to identify all such method calls between classes and obtain a set of edges, $E$ between the corresponding nodes. The edges are unweighted and multiple method calls from class A to class B are still represented by a single edge from A to B.


We now describe the process to generate the attribute matrix, $X$, coressponding to the nodes $V$ of the graph. Most modern web applications expose multiple APIs that perform various functions. These APIs (UI elements in the case of a non web-based application) are referred to as EntryPoint Specifications \cite{dietrich2018driver}, or simply, \textit{Entrypoints}. The methods invoked through these APIs are specially annotated as such and are called \textit{entrypoint methods} in this work. Figure \ref{fig:code_graph} shows an example of such entrypoint methods annotated with \textit{@API}. We refer to the classes containing such entrypoint methods as \textit{entrypoint classes}. Each entrypoint class can thus be associated with multiple Entrypoints. 
Starting from an entrypoint method, we can follow the call sequence of methods through the application, keeping track of all classes invoked during the execution trace of that Entrypoint. If $P$ is the set of Entrypoints in an application, we can define a matrix $EP^{|V| \times |P|}$, such that $EP(i,p) = 1$ if class $i$ is present in the execution trace of entrypoint $p$, else $0$. Additionally, we define $C^{|V| \times |V|}$ such that $C(i,j)$ is the number of Entrypoint execution traces that contain both classes $i$ and $j$. If a class is not invoked in an execution trace for any Entrypoint, we remove the corresponding node from the graph. Finally, classes may also inherit from other classes or Interfaces. In Figure \ref{fig:code_graph}, class \textit{A} inherits from class \textit{Base}. Although this establishes a dependency between the classes, it does not involve direct method invocation. Hence, we do not include this dependency as an edge in the graph, but as a node attribute. Therefore, we set $In(i,j) = In(j,i) = 1$ if classes $i$ and $j$ are related via an inheritance relationship and $0$ otherwise.
The attribute matrix $X$ is the concatenation of $EP$, $C$ and $In$ matrices. Thus, $X \in \mathbb{R}^{|V| \times F}$ where $F=|P|+2|V|$. Each constituent of $X$ is row-normalized individually. The application can thus be represented as a graph $G=(V,E,X)$.

\subsection{Proposed Graph Neural Network}\label{sec:GNN}
Given the graph $G$, we want to develop a graph neural network which can jointly (i) derive vector representations (embeddings) of the nodes, (ii) minimize the effect of outlier nodes in the embeddings of other regular nodes, (iii) obtain communities in the graph.
Let us use $A \in \mathbb{R}^{|V| \times |V|}$ to denote the adjacency matrix of $G$, where $A_{ij}$ is the weight of the edge $e_{ij}$ if it exists, otherwise $A_{ij}=0$. We use a 2-layered graph convolution encoder \cite{kipf2017semi} to obtain representation of each node as shown below:
\begin{align}\label{eq:encoder}
    Z = f(X, A) = \text{ReLU}(\hat{A} \; \text{ReLU}(\hat{A}XW^{(0)}) \; W^{(1)})
\end{align}
where each row of $Z \in \mathbb{R}^{|V| \times F'}$ contains the corresponding node representation. We compute $\Tilde{A} = A + I$, where $I \in \mathbb{R}^{|V| \times |V|}$ is the identity matrix and the degree diagonal matrix $\Tilde{D_{ii}}$ with $\Tilde{D_{ii}} = \sum\limits_{j \in V} \Tilde{A}_{ij}$, $\forall i \in V$. We set $\hat{A} = \Tilde{D}^{-\frac{1}{2}} \Tilde{A} \Tilde{D}^{-\frac{1}{2}}$.
$W^{(0)}$ and $W^{(1)}$ are the trainable parameter matrices of GCN encoder.
Traditionally, these parameters are trained on a node classification or link prediction loss \cite{kipf2016variational} in a graph.

However, our objective in the work is to consider and minimize the effect of outlier nodes in the framework of graph convolution. We also want to do this in an unsupervised way as obtaining ground truth class labels and outlier information are extremely difficult for monolith applications. Towards this, we use the following GCN based decoder to map the $F'$ dimensional node embeddings to the input feature space.
\begin{align}\label{eq:decoder}
    \hat{X} = f(Z, A) = \text{ReLU}(\hat{A} \; \text{ReLU}(\hat{A}ZW^{(2)}) \; W^{(3)})
\end{align}

Here, $\hat{X} \in \mathbb{R}^F$, $W^{(2)}$ and $W^{(3)}$ are the trainable parameters of the decoder. Let us use $\mathcal{W} = \{W^{(0)},\cdots,W^{(3)}\}$ to denote the parameters of the encoder and decoder combined. In the ideal world scenario when there is no outlier node present in a graph, one can train the parameters of the GCN autoencoder by directly minimizing some reconstruction loss. But as mentioned in Section \ref{sec:intro}, the presence of outliers in monolith applications is prevalent and if not handled properly, they can adversely affect the embeddings of regular regular nodes in a graph \cite{bandyopadhyay2020outlier}.
To address them, we use the framework of multi-task learning where we design two loss components to detect structural and attribute outliers respectively. We denote structural and attribute outlierness (positive scalars) by $O_{si}$ and $O_{ai}$ respectively, for each node $i \in V$.

First, we ensure that presence of an edge should be preserved by the similarity of the two corresponding node embeddings in the vector space for the regular nodes. However, structural outliers being inconsistent in their link structure, do not necessarily follow this assumption. Hence, we design the following loss component which needs to be minimized with respect to the parameters of GCN and structural outlierness of the nodes:
\begin{align}\label{eq:loss1}
    \mathcal{L}_{str} = \sum\limits_{i \in V} \log \Big(\frac{1}{O_{si}}\Big) || A_{i:} - Z Z_{i:}^T ||_2^2
\end{align}
Here, $A_{i:}$ is the $i$th row of the adjacency matrix and $Z_{i:}$ is the $i$th row (embedding of node $i$) of the node representation matrix. Clearly, higher the value of $O_{si}$, i.e., higher the outlierness of node $i$, less will be the value of $\log \Big(\frac{1}{O_{si}}\Big)$. Consequently, contribution of the structural outlier nodes in this loss component will be less. 
We also assume that total structural outlierness in a graph is bounded. So we set $\sum\limits_{i \in V} O_{si} = 1$. Without such a bound, the optimization in Equation \ref{eq:loss1} would reach to a degenerate solution with each $O_{si}$ assigned to $+\infty$ at the infimum. We also tried replacing $1$ with a hyperparameter $\mu$ as the bound, but that does not have much impact on the quality of the final solution.

Next, to preserve the impact of node attributes in the node representations, we want the reconstructed attributes in Equation \ref{eq:decoder} from the GCN decoder to match the initial node attributes for most of the regular nodes in the graph. However for attribute outliers, as their node attributes are significantly different from the attributes of their respective neighboring nodes, we reduce their contribution in the attribute reconstruction loss as follows:
\begin{align}\label{eq:loss2}
    \mathcal{L}_{att} = \sum\limits_{i \in V} \log \Big(\frac{1}{O_{ai}}\Big) || X_{i:} - \hat{X}_{i:} ||_2^2
\end{align}

Here, $X$ and $\hat{X}$ are the given and reconstructed node feature matrices. Similar to the case of structural outlierness, nodes with more attribute outlierness score $O_{ai}$ would have less impact in Equation \ref{eq:loss2} and consequently the optimizer will be able to focus more on the regular nodes of the graph. Again, we assume that $O_{ai} > 0$, $\forall i \in V$ and $\sum\limits_{i \in V} O_{ai} = 1$.

Minimizing the loss components in Equations \ref{eq:loss1} and \ref{eq:loss2} with respect to the parameters of GCN and outlier scores would be able to provide unsupervised node embeddings. This will also detect the outlier nodes while minimize their negative impact on the other nodes of the graph. However as discussed in Section \ref{sec:intro}, our main goal in this work is to separate microservices within a monolith application. This needs discovering clusters of nodes (or communities) in the graph. One can potentially obtain the node embeddings first and then use a clustering algorithm (for example, k-means++ \cite{arthur2006k}) as a post-processing step. But such a decoupled approach often leads to a suboptimal solution as shown in \cite{yang2017towards}. Hence, we integrate node embedding, outlier detection and node clustering in a joint framework of graph neural network. To achieve this, we use the following loss to cluster the nodes in the graph, assuming their embeddings are already given.
\begin{align}\label{eq:loss3}
	\mathcal{L}_{clus} = \sum\limits_{i=1}^N \sum\limits_{k=1}^K M_{ik} || Z_{i:} - C_k ||_2^2
\end{align}
where $M \in \{0,1\}^{|V| \times K}$ is the binary cluster assignment matrix. We assume to know the number of clusters $K$. $M_{ik} = 1$ if node $i$ belongs to $k$th cluster and $M_{ik} = 0$ otherwise. $C_k \in \mathbb{R}^{F'}$ is the center of each cluster in the embedding space. Equation \ref{eq:loss3} needs to be minimized with respect to $M$ and $C = [C_1 \cdots C_K]^T$ to obtain the clustering.
We call this method CO-GCN (\underline{C}lustering and \underline{O}utlier aware \underline{G}raph \underline{C}onvolution \underline{N}etwork) and the joint loss function is:
\begin{align}\label{eq:totalLoss}
\underset{\mathcal{W},\mathcal{O},M,C}{\text{min}} \; & \mathcal{L}_{total} = \alpha_1 \mathcal{L}_{str} + \alpha_2 \mathcal{L}_{att} + \alpha_3 \mathcal{L}_{clus} \\
\text{such that, }\;\; & \sum\limits_{i \in V} O_{si} = \sum\limits_{i \in V} O_{ai} = 1 \\
& M \in \{0,1\}^{|V| \times K}, \;\; O_{si}, O_{ai} > 0 \;\; \forall i \in V
\end{align}

\subsection{Optimization Procedure}\label{sec:opti}
The nature of the optimization problem in Eq. \ref{eq:totalLoss} is different with respect to different variables. We use alternate minimization technique, where we minimize the objective only with respect to one set of variables, keeping others fixed.

\subsubsection{Parameters of GCN}
The set $\mathcal{W}$ contains all the parameters of the GCN encoder and decoder as described in Section \ref{sec:soln}. We use standard ADAM optimization technique \cite{kingma2014adam} to minimize the total loss w.r.t. $\mathcal{W}$, keeping other variables fixed. We use an initial learning rate of $0.01$ and exponential decay rate of $0.95$ every $100$ iterations.

\subsubsection{Outliers} One can show that optimization in Equation \ref{eq:totalLoss} is convex with respect to each outlier variable when all other variables are fixed. This is because $0 < O_{si}, O_{ai}  \leq 1$, $\forall i$ and $\log(\cdot)$ is a concave function and thus, $-\log(\cdot)$ is convex. Finally, L2 norms in both Equations \ref{eq:loss1} and \ref{eq:loss2} are non-negative. We aim to find the closed form update rules for the outlier terms to speed up the optimization process.

Taking the Lagrangian of Eq. \ref{eq:totalLoss} with respect to the constraint $\sum\limits_{i \in V} O_{si} = 1$, we get (after ignoring terms that do not include $O_{si}$),
\begin{flalign}
& \frac{\partial }{\partial O_{si}} \sum\limits_{j \in V}\log\Big({\frac{1}{O_{sj}}}\Big) || A_{j:} - Z_{j:}^T Z ||_2^2 + \lambda(\sum\limits_{j \in V} O_{sj} - 1) \nonumber  
\end{flalign}
$\lambda \in \mathbb{R}$ is the Lagrangian constant. Equating the partial derivative w.r.t. $O_{si}$ to 0:
\begin{align*}
&-\frac{|| A_{i:} - Z_{i:}^T Z ||_2^2}{O_{si}} + \lambda = 0, \;
\Rightarrow O_{si} = \frac{|| A_{i:} - Z_{i:}^T Z ||_2^2}{\lambda}
\end{align*}
But, $\sum\limits_{j=1}^N O_{ji} = 1$ implies $\sum\limits_{j \in V} \frac{|| A_{j:} - Z_{j:}^T Z ||_2^2}{\lambda} = 1$. Hence,
\begin{align}\label{eq:UpdateOa}
O_{si} = \frac{|| A_{i:} - Z_{i:}^T Z ||_2^2}
{
\sum\limits_{j \in V} || A_{j:} - Z_{j:}^T Z ||_2^2
}
\end{align}
The final update rule for structural outliers turns out to be quite intuitive. Our goal while deriving the loss in Equation \ref{eq:loss1} was to approximate adjacency structure of the graph by the similarity in the embedding space with outliers being discounted. The structural outlierness of a node in Equation \ref{eq:UpdateOs} is proportional to the difference between the two after every iteration. In other words, if some node is not able to preserve its adjacency structure in the embedding space, it is more prone to be a structural outlier.

Similar to above, update rule for attribute outlier at each iteration can be derived to the following.
\begin{align}\label{eq:UpdateOs}
O_{ai} = \frac{|| X_{i:} - \hat{X}_{i:} ||_2^2}
{
\sum\limits_{j \in V} || X_{j:} - \hat{X}_{j:} ||_2^2
}
\end{align}

Because of the convexity of total loss in Equation \ref{eq:totalLoss} w.r.t. individual outlier scores, derivations of the update rules for outlier scores ensure the following lemma.
\begin{lemma}\label{lemma:1}
Keeping all other variables fixed, the total loss in Equation \ref{eq:totalLoss} decreases after every update of the outlier scores by Equations \ref{eq:UpdateOs} and \ref{eq:UpdateOa} until it reaches to a stationary point.
\end{lemma}

\subsubsection{Clustering Parameters}
The total loss of CO-GCN also involves clustering parameters $M$ and $C$. While all other variables to be fixed, cluster assignment matrix $M$ can be obtained as:
\begin{flalign}\label{eq:clusAssign}
M(i,k)=
\begin{cases}
1, \; \text{if\;} k = \underset{k' \in \{1,\cdots,K\}}{\text{argmin}} ||Z_i - C_{k'} ||_2^2\\
0, \; \text{Otherwise}
\end{cases}
\end{flalign}
In the next step, $k$th row of cluster center matrix $C$ can be obtained as \cite{arthur2006k}:
\begin{flalign}\label{eq:clusCen}
C_{k} = \frac{1}{N_k} \sum\limits_{i \in \mathcal{C}_k} Z_{i:} 
\end{flalign}
where $\mathcal{C}_k = \{i \in V \;|\; M_{ik}=1 \}$ is the $k$-th cluster and $N_k = |\mathcal{C}_k|$ is the size of $k$-th cluster.

\subsection{Pre-training, Algorithm and Analysis}\label{sec:alg}
To run CO-GCN, we first pre-train the GCN encoder and decoder by minimizing $\mathcal{L}_{str}$ and $\mathcal{L}_{att}$ in Equations \ref{eq:loss1} and \ref{eq:loss2} respectively, initializing $O_{si}$, $O_{ai}$ $\forall i \in V$ to uniform values. We also use k-means++ \cite{arthur2006k} to initialize the cluster assignment and cluster center matrices. Then over iterations, we sequentially solve $\mathcal{L}_{total}$ by alternating minimization technique described in Section \ref{sec:opti} with respect to different variables. Overall procedure of CO-GCN is presented in Algorithm \ref{alg:cogcn}.

\begin{algorithm} 
  \small
  \caption{\textbf{CO-GCN}}
  \label{alg:cogcn}
\begin{algorithmic}[1]
      
	\Statex \textbf{Input}: Class dependencies and Entrypoint definitions 
	\State Convert the application to a graph representation as defined in Section \ref{sec:convGraph} and obtain the $V$, $E$ and $X$
    \State Initialize outlier scores $O_{si}$ and $O_{ai}$ uniformly $\forall i \in V$.
    \State Pre-train the GCN encoder and decoder
    \State Use k-means++ to initialize the cluster assignment and cluster centers
	\For{T iterations}
    	\State Update outlier scores $\mathcal{{O}}$ by Eq. \ref{eq:UpdateOs} and \ref{eq:UpdateOa}.
		\State Update cluster assignment and center matrices by Eq. \ref{eq:clusAssign} and \ref{eq:clusCen}
        \State Update the parameters by GCN encoder and decoder by minimizing Eq. \ref{eq:totalLoss} using ADAM.
    \EndFor
    \Statex \textbf{Output}: Cluster assignment matrix $M$, Cluster center matrix $C$ and the outlier scores $\mathcal{O}$
	\end{algorithmic}
  \end{algorithm} 

\subsubsection{Time Complexity} Time taken by GCN encoder and decoder is $O(|E|FF')$. Updating each value of outlier score takes $O(NF')$ and the total time to update all outlier scores is $O(N^2F')$. Updating the parameters of cluster assignment and cluster center matrices takes $O(NF'K)$ time. Thus, each iteration of CO-GCN takes $O(|E|FF' + N^2F' + NF'K)$. The outlier update rules although expensive, converge quickly because of the closed-form solution and theoretical guarantee (Lemma \ref{lemma:1}). Also, for most real-world monolith applications, number of classes is not very large (in 1000s). So the quadratic dependency of the runtime on the number of classes is not a bottleneck. However, one can try negative sampling approaches \cite{goldberg2014word2vec} to approximate the similarity between the embeddings in the outlier update rules for other applications if needed.


\section{Experimental Evaluation}
\label{sec:eval}

\begin{table}[t]
\centering
\resizebox{\linewidth}{!}{%
\begin{tabular}{|l|l|l|l|l|l|l|}
\hline
\textbf{Dataset} & \textbf{Description} & \textbf{Language} & \textbf{\# Class} & \textbf{\# Entrypoints} & \textbf{Cluster size} \\
\hline
DayTrader & Trading App & Java & 111  & 203 & 8 \\
\hline
PBW  & Online plant store & Java & 36  & 47 & 6 \\
\hline
Acme-Air & Airline App & Java & 38  & 20 & 4 \\
\hline
DietApp  & DietTracker & C\# & 32 & 34 & 5 \\
\hline
\end{tabular}
}
\caption{Details about the monolith applications studied}
\label{tab:dataset}
\end{table}

\subsection{Datasets (Monolith Applications) Used}
 To study the effectiveness of our approach, we chose four publicly-available web-based monolith applications namely Daytrader \footnote{https://github.com/WASdev/sample.daytrader7}, Plantsbywebsphere \footnote{https://github.com/WASdev/sample.plantsbywebsphere}, Acme-Air\footnote{https://github.com/acmeair/acmeair}, Diet App\footnote{https://github.com/SebastianBienert/DietApp/}.
They vary in programming languages, technologies, objectives and complexity in terms of lines of code, function sizes etc.
Details of the monoliths are provided in Table \ref{tab:dataset}.


\subsection{Metrics}
\label{sec:metrics}


\begin{figure*}[!t]
  \centering
  \includegraphics[width=0.9\textwidth]{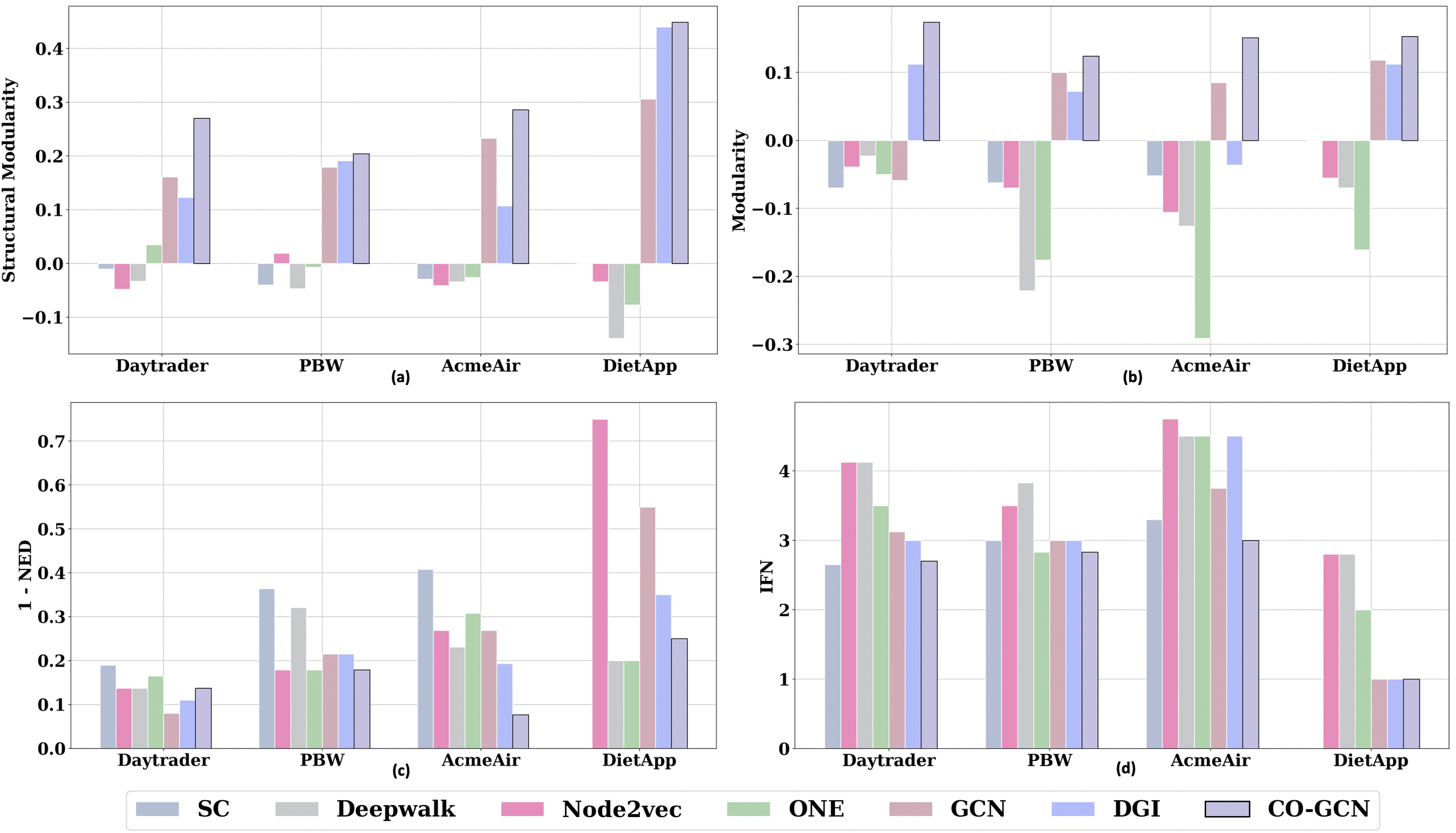}
  \caption{Comparison of the CO-GCN method with the baselines across the four applications on the (a) Structural Modularity (b) Modularity (c) 1-NED and (d) IFN metrics. The CO-GCN method clearly outperforms the baselines considered.}
  \label{fig:results}
\end{figure*}

To evaluate the quality of the clusters identified as microservice candidates, we define four metrics. The first two aim to capture the structural quality of the clusters recommended as microservices and are the primary metrics in the evaluation. The other two metrics define additional properties of the clusters that are desirable.

\begin{enumerate}
    \item \textbf{Modularity}:
    Modularity is a commonly used metric to evaluate the quality of clusters in a graph \cite{modularity1}\cite{modularity2}. It measures the fraction of edges of the graph between members of the same cluster relative to that of the same partition members but randomly generated graph edges. Higher values of Modularity indicate a stronger community structure.
    
    \item \textbf{Structural Modularity}:
    An alternate measure of structural soundness of a cluster that is more suited to software applications is defined in \cite{jin2019service}. Structural Modularity, (SM) is defined as 
    \begin{equation*}
      SM = \frac{1}{K}\sum \limits_{k=1}^K \frac{u_k}{N^2_k}  - \frac{1}{K(K-1)/2}\sum \limits_{k_1 \not=k_2}^K \frac{\sigma_{k_1,k_2}}{2(N_{k_1} N_{k_2})}
    \end{equation*}
    and $u_k$ is the number of edges that lie completely within a cluster $k$, $\sigma_{k_1,k_2}$ is the number of edges between cluster $k_1$ and cluster $k_2$. $N_{k_1}$ and $N_{k_2}$ are the number of members in clusters $k_1$ and $k_2$ respectively.
    
    \item \textbf{Non-Extreme Distribution(NED)}: 
    It is desired that a microservice may not have too many or too few classes. We therefore measure how evenly distributed the sizes of the recommended clusters are as
    \begin{equation*}
      NED = \frac{\sum^K_{k=1,k\:not\:extreme}n_k}{|V|}
        \end{equation*}
    $n_k$ is the number of classes in cluster $k$ and $V$ is the set of classes. $k$ is not extreme if it's size is within bounds of $\{5,20\}$. $NED$ captures the architectural soundness of the clusters \cite{NED1}\cite{NED2}. For better interpretability, we measure $1 - NED$ and lower values are favorable.
    
    \item \textbf{Interface Number(IFN)}:
    As defined in \cite{jin2019service}, this is the average number of published interfaces of a microservices partitioning. 
    \begin{equation*}
      IFN = \frac{1}{K}\sum \limits_{k=1}^K ifn_k , \;   \;  ifn_k = |I_k|
    \end{equation*}
    
    where $I_k$ is the number of published interfaces in the microservice $k$ and $K$ is the number of such micorservices. We define a published interface as any class in the microservice that is referenced by another class from a different microservice. Lower values of $IFN$ are preferred.
\end{enumerate}

\subsection{Experimental Setup and Baselines}
For each application in Table \ref{tab:dataset}, we generate the adjacency matrix, $A$ and the feature matrix, $X$. The CO-GCN encoder comprises of two layers with dimensionality $64$ and $32$. The decoder consists of one layer of size $64$ followed by another of the appropriate feature dimension. We pretrain for $250$ iterations and set $T=500$ in Algorithm \ref{alg:cogcn}.
The final values of $M(i,k)$ are used as the cluster assignments from our algorithm. 
We set $\{\alpha_1, \alpha_2, \alpha_3 \} = \{0.1, 0.1, 0.8\}$ in Eq. \ref{eq:totalLoss}. 

We evaluate our approach against multiple unsupervised baselines for learning node representations: Deepwalk \cite{perozzi2014deepwalk},  Node2vec \cite{grover2016node2vec}, ONE \cite{bandyopadhyay2019outlier}
GCN \cite{kipf2016variational} and DGI
\cite{velivckovic2018deep}. Among these, ONE accounts for the effects of outliers in learning node embeddings. For all our experiments, we set the size of the node embeddings to be $32$. We use k-means++ algorithm on the embeddings generated by these baselines to obtain clusters. K is carefully chosen based on online sources and SME inputs.
In contrast to these representation learning based baselines, the method of \cite{mazlami-icws-2017} is a state-of-the-art approach for extracting microservices from a monolith application. This leverages Semantic Coupling (SC) information with graph partitioning to identify clusters. We also use it as a baseline.
Since the implementation for the SC method does not support .Net applications, we do not use it for DietApp.




\begin{figure}[!t]
  \centering
  \includegraphics[width=0.45\textwidth]{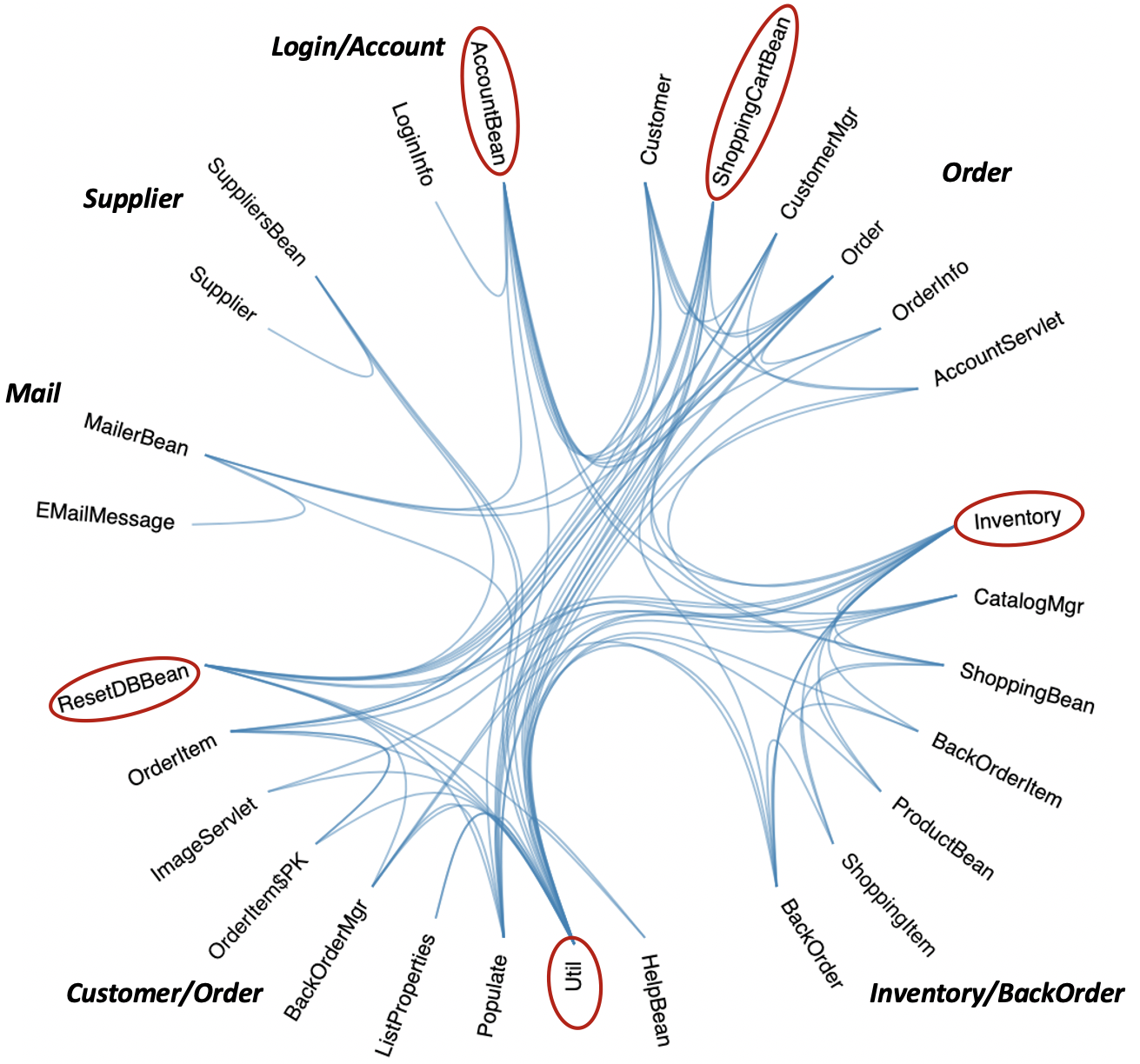}
  \caption{Clusters and top 5 outliers identified for the PBW application, with manual labels about their functionality.}
  \label{fig:pbwclusters}
\end{figure}

\begin{figure}[!t]
  \centering
  \includegraphics[width=\linewidth]{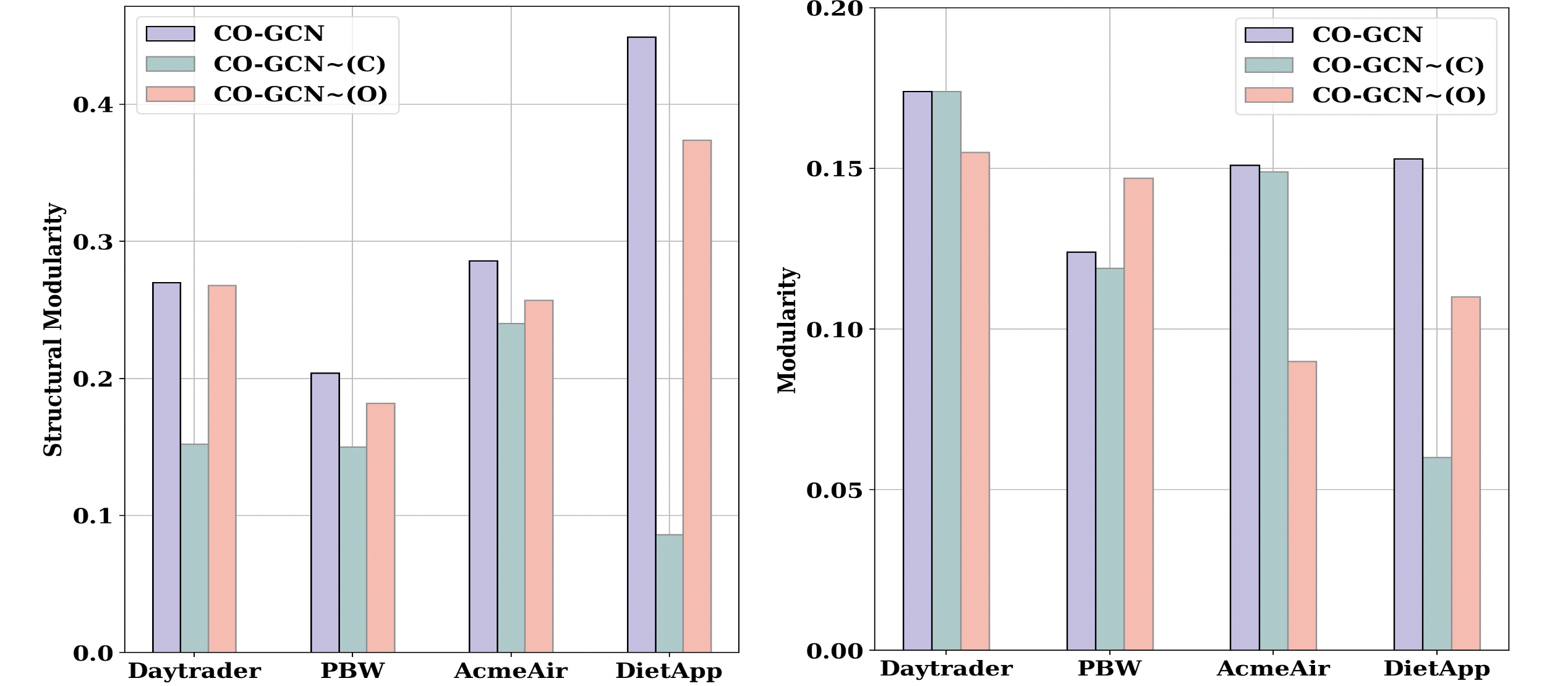}
  \caption{Results from the ablation study on the structural modularity and modularity metrics across the applications}
  \label{fig:ablation}
\end{figure}


\subsection{Results on Separating Micro Services}
Figure \ref{fig:results} shows the metrics values on all four application for the evaluated methods. The three attributed graph neural network based methods (GCN, DGI and CO-GCN) outperform the rest of the methods by a significant margin. The CO-GCN method consistently achieves better modularity and structural modularity scores which clearly validates the inclusion of outlier and clustering objectives in the training. The CO-GCN method also achieves better $NED$ and $IFN$ scores in most cases. 
Another interesting observation is the negative scores for many of the baseline methods. This implies that are many inter-cluster edges for the clusters recommended by these methods, hinting at the fact that monolith applications may have several high-traffic nodes and assigning them to appropriate clusters is difficult, but critical. Figure \ref{fig:pbwclusters} shows the identified clusters for the PBW application and our manual annotations to highlight the functionalities offered. We can notice the clear distinction of functionalities 


\subsection{Detecting Outliers}
The values of $O_{si}$ and $O_{ai}$ at the end of training represent the final outlier scores of each node. The ranked list of outlier nodes represents the top candidates for refactoring as part of microservices decomposition. Figure \ref{fig:pbwclusters} highlights the combined top $5$ outliers detected (across structural and attribute outlier scores) for PBW application by our approach. 
Among the baselines, we report outlier detection results only for GCN and DGI as they performed good for obtaining microservices. As GCN and DGI do not output outliers directly, we use Isolation forest \cite{liu2008isolation} on the embeddings generated by them to detect outliers. 


To study the goodness of the outliers, we performed a qualitative study with five software engineers who have minimum seven years industrial experience. We randomly presented them with two out of the four monoliths and shared their code repositories. We asked them to rank the top five refactor candidate classes and compared them with the outliers identified by GCN, DGI and CO-GCN. On an average, the top five outliers provided by the annotators overlapped with our approach by 60\%, GCN by 45\% and DGI by 55\%. Details of top $5$ outliers detected by each approach, our questionnaire and the results are provided in the supplementary material. We can conclude that the outliers identified by our approach are more relevant. The low overlap numbers indicate the highly difficult and subjective nature of this task.

\subsection{Ablation and Sensitivity Analysis}
We perform another set of experiments to measure the usefulness of individual components of CO-GCN.
\begin{enumerate}
    \item We remove the clustering objective from $\mathcal{L}_{total}$., i.e., set $\alpha_3=0$ in Equation \ref{eq:totalLoss}.
    Comparing the performance of this variant with CO-GCN shows marginal contribution of integrating the clustering loss. We denote this variant as CO-GCN\texttildelow(C). We use k-means++ on the node embeddings generated by this approach to obtain the clusters.
    \item We remove the effect of the $O_{si}$ and $O_{ai}$ on $\mathcal{L}_{str}$ and $\mathcal{L}_{att}$ respectively, by removing the $\log(\cdot)$ terms. This is equivalent to traditional link and attribute reconstruction, with the clustering loss $\mathcal{L}_{clus}$.
    The goal is to evaluate the usefulness of minimizing the effect of outliers for identifying good clusters. We denote this variant as CO-GCN\texttildelow(O).
\end{enumerate}

The results of the ablation study are shown in Figure \ref{fig:ablation}. In general, incorporating outlier scores and the clustering objective does result in higher modularity and structural modularity scores. However, the degree to which these components contribute to the overall clustering quality vary for each application and the metric used. For instance, in the Daytrader application, removing the clustering objective reduces structural modularity significantly, but has no effect on modularity. Conversely, removing the outlier information reduces the modularity score, but has negligible effect on structural modularity. This effect is also visible in the other applications. 
Interestingly, removing the outlier information leads to improved modularity for PBW, but this is balanced by a reduced structural modularity score. We can still conclude that including the outlier scores and clustering loss in the training objective improves cluster quality in general.

\begin{figure}[!t]
  \centering
  \includegraphics[width=\linewidth]{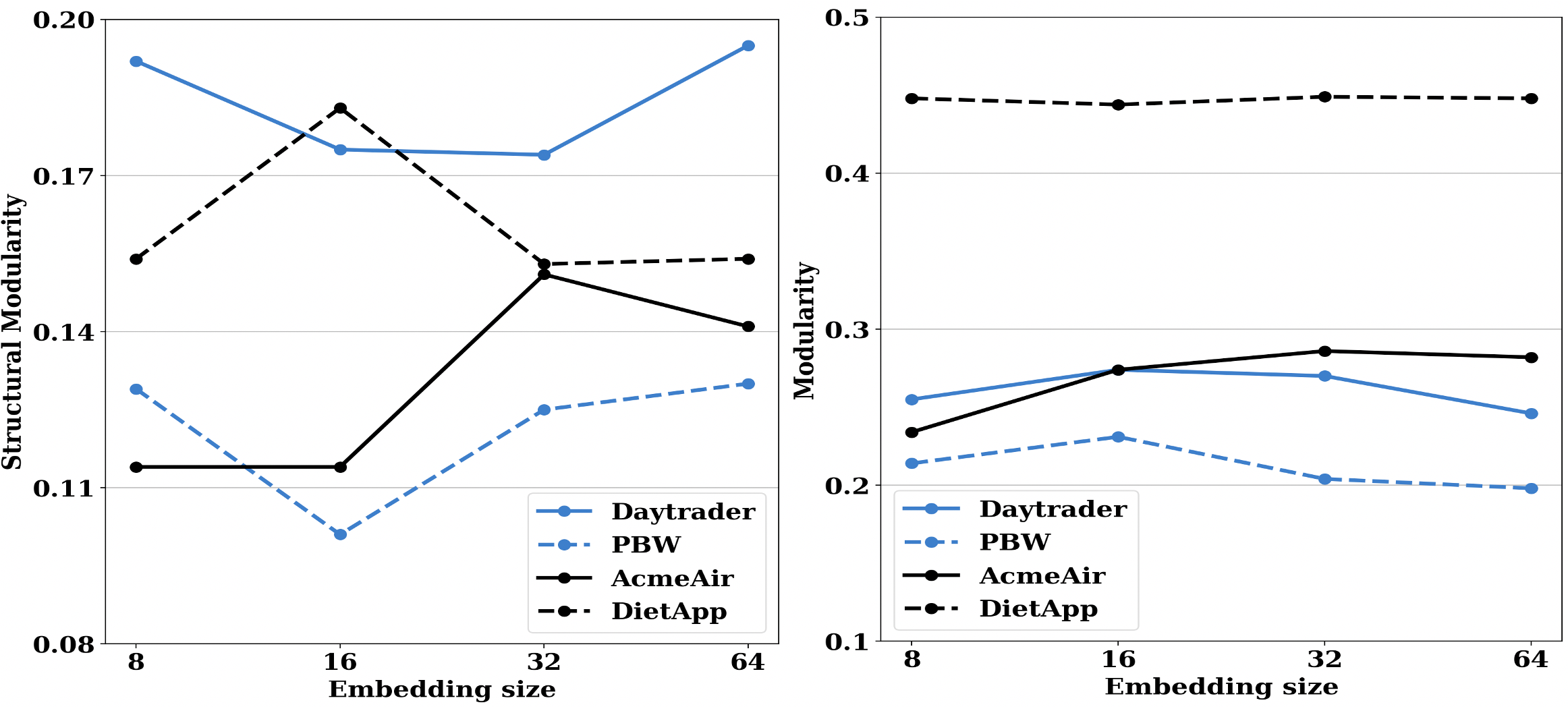}
  \caption{Sensitivity analysis on embedding size}
  \label{fig:sensiplot}
\end{figure}

Finally, we also evaluate the effect of the node embeddings size on the modularity and structural modularity values for each application. We experiment with embedding sizes in $\{8,16,32,64\}$. The results are presented in Figure \ref{fig:sensiplot}. We notice the modularity scores do not have any significant variation with a change in node embedding size. There is relatively more variation in the structural modularity scores with change in embedding sizes and once again, this variation is application dependent. There is not enough evidence to make any substantial claims, but in general, the performance seems to be better at higher embedding sizes.

\section{Conclusion}\label{sec:conclusion}

We introduced the traditional software engineering problem of monolith to microservices decomposition as a clustering task building upon graph representation learning. We showed how the application implementation structure can be translated into an attributed graph network. We then proposed a multi-objective Graph Convolution Network (GCN) based novel framework to not just generate clusters which can be candidate microservices, but also identified the outliers in the graph which can be considered as the important refactor classes for the architect to focus on. Our approach improved state of the art on multiple metrics from both graph and software engineering literature and performed better than others in human evaluation for the outlier detection. In future, we want to extend this work to automatically identify the number of microservices and expand the studies to procedural programming languages like COBOL.

\section{Potential Ethical Impact (Non-Technical)}
We believe this work doesn't have any direct societal or ethical impact. 

\section{Acknowledgements}
We would like to thank Giriprasad Sridhara, Amith Singhee, Shivali Agarwal, Raunak Sinha from IBM India Research Labs and Yasu Kastuno, Ai Ishida, Aki Tozawa, Fumiko Satoh from IBM Tokyo Research Labs for their insightful suggestions during this work and valuable feedback towards improving the paper.

\bibliography{references}

\clearpage
\newpage
\appendix


\section{Notation}

Different notations used in the paper are summarized in Ta-2ble 1

\section{Outliers}\label{sec:outlier}

\subsection{Examples of an outlier from the source code}
\begin{figure*}[!t]
  \centering
  \includegraphics[width=\linewidth]{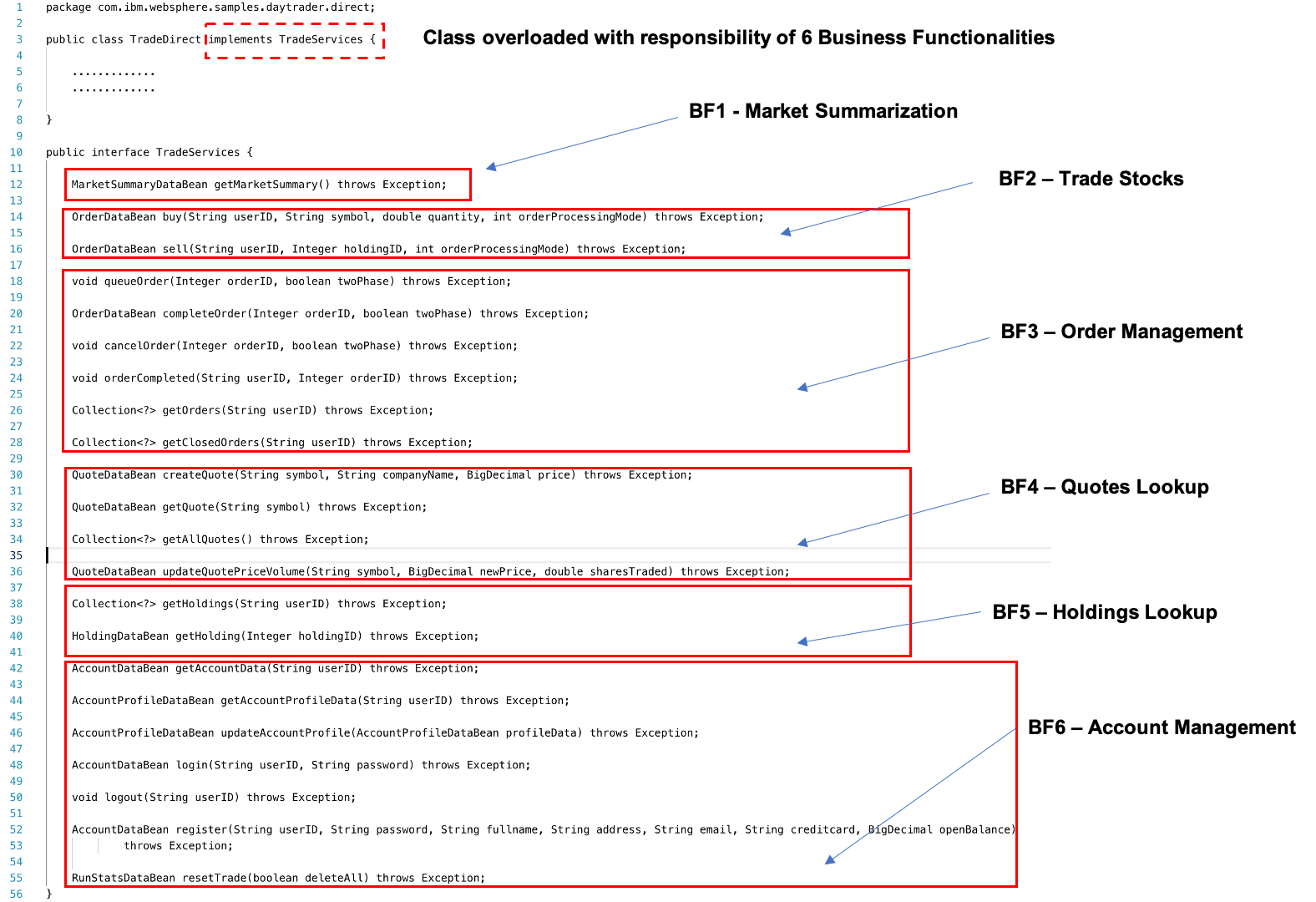}
  \caption{TradeDirect class implements TradeServices interface to support six different business functionalities through the twenty two methods. This overladed business functionality classs is considered outlier class and becomes a candidate refactorable class for the developer to break into six classes containing only specific business function methods}
  \label{fig:daytrader_outlier}
\end{figure*}

We define outliers as classes that have no unique business functionality identity. Primarily, these are classes that are  overloaded with multiple business functionalities as seen in Figure \ref{fig:daytrader_outlier}. \textit{Daytrader \footnote{https://github.com/WASdev/sample.daytrader7} is an online stock trading monolith application. The application allows users to setup the trading platform like configuring the database, user account and trading amount etc. Once setup the platform allows users to view market summary, check stock quotes, and buy or sell stocks.} TradeDirect is one of the core class in the application which implements TradeServices interface. This interface exposes six different abstract functionalities through roughly twenty methods to implement. TradeDirect implements the interface and provides logical support to all six business functionalities. Now, the identity of this class is overloaded and therefore, become top refactor candidates for the developers. Developers tend to breakdown the methods and separate them into six classes specific to the business functionality. There can be other type of outliers too like classes that doesn't support any business functionality but the core business function classes depend on them for completion. They can be identified by dependencies from multiple classes. Typically, utilities display such a behavior. \ref{tab:outliers} lists the top five outliers determined by the three attributed graph neural network based methods (CO-GCN, GCN and DGI) for each of the four monolith applications namely Daytrader \footnote{https://github.com/WASdev/sample.daytrader7}, Plantsbywebsphere (PBW) \footnote{https://github.com/WASdev/sample.plantsbywebsphere}, Acme-Air\footnote{https://github.com/acmeair/acmeair}, Diet App\footnote{https://github.com/SebastianBienert/DietApp/}.

\subsection{Details of Outlier Evaluation Study}
To study the goodness of the outliers, we performed a qualitative study with five software engineers who have minimum seven years industrial experience. We randomly presented them an overview of two out of the four four monoliths and shared their code repositories. We requested them to manually go through the program classes to understand the functionality exposed and the classes that come together to support them. We intentionally did not explain the application details since it might influence their perspective of outliers. Application understanding is an time consuming and specialized activity. All the annotators had prior experience in building web applications and are proficient in the language used in the monolith applications. All the annotator took roughly twenty four hours to study the monoliths. 
For the study, we asked each annotator to rank the top five refactor candidate classes for the two application they were presented. We then compared them with the outliers identified by CO-GCN, GCN and DGI methods. On an average, the top five outliers provided by the annotators overlapped with our approach by 60\%, GCN by 45\% and DGI by 55\%. Details of the overlap for the annotators for each of the approaches for the four applications are listed in Table \ref{tab:daytrader_outliers_study}, Table \ref{tab:pbw_outliers_study} ,Table \ref{tab:acmeair_outliers_study} and Table \ref{tab:dietapp_outliers_study}. We can conclude that the outliers identified by our approach are more relevant. The low overlap numbers also indicate the highly difficult and subjective nature of this task.

\begin{table*}[t]
\centering
\begin{tabular}{|c|c|c|}\hline
\textbf{Dataset} & \textbf{Approaches} & \textbf{Top 5 Outliers classes} \\
\hline
\multirow{15}{*}{Daytrader} & \multirow{5}{*}{CO-GCN}  & 1. com.ibm.websphere.samples.daytrader.web.jsf.AccountDataJSF \\\cline{3-3}
& & 2.	com.ibm.websphere.samples.daytrader.web.prims.PingJSONP \\\cline{3-3}
& & 3.	com.ibm.websphere.samples.daytrader.util.FinancialUtils \\\cline{3-3}
& & 4.	com.ibm.websphere.samples.daytrader.direct.TradeDirect \\\cline{3-3}
& & 5.	om.ibm.websphere.samples.daytrader.TradeAction \\\cline{2-3}

& \multirow{5}{*}{GCN}  & 1.	com.ibm.websphere.samples.daytrader.web.prims.PingCDIBean \\\cline{3-3}
& & 2.	com.ibm.websphere.samples.daytrader.util.TradeConfig \\\cline{3-3}
& & 3.	com.ibm.websphere.samples.daytrader.util.Log \\\cline{3-3}
& & 4.	com.ibm.websphere.samples.daytrader.util.TradeDirect \\\cline{3-3}
& & 5.	com.ibm.websphere.samples.daytrader.web.prims.PingJSONP \\\cline{2-3}

& \multirow{5}{*}{DGI}  & 1. com.ibm.websphere.samples.daytrader.web.jsf.QuoteData \\\cline{3-3}
& & 2.	com.ibm.websphere.samples.daytrader.util.KeyBlock \\\cline{3-3}
& & 3.	com.ibm.websphere.samples.daytrader.util.Log \\\cline{3-3}
& & 4.	com.ibm.websphere.samples.daytrader.web.prims.PingUpgradeServlet \\\cline{3-3}
& & 5.	com.ibm.websphere.samples.daytrader.web.prims.PingServletCDI \\
\hline

\multirow{15}{*}{PBW} & \multirow{5}{*}{CO-GCN}  & 1. com.ibm.websphere.samples.pbw.ejb.ResetDBBean \\\cline{3-3}
& & 2.	com.ibm.websphere.samples.pbw.war.ShoppingBean \\\cline{3-3}
& & 3.	com.ibm.websphere.samples.pbw.war.AccountBean \\\cline{3-3}
& & 4.	com.ibm.websphere.samples.pbw.ejb.CustomerMgr \\\cline{3-3}
& & 5.	com.ibm.websphere.samples.pbw.jpa.OrderItem \\\cline{2-3}

& \multirow{5}{*}{GCN}  & 1.	com.ibm.websphere.samples.pbw.ejb.EMailMessage \\\cline{3-3}
& & 2.	com.ibm.websphere.samples.pbw.ejb.MailerBean \\\cline{3-3}
& & 3.	com.ibm.websphere.samples.pbw.war.AccountBean \\\cline{3-3}
& & 4.	com.ibm.websphere.samples.pbw.war.LoginInfo \\\cline{3-3}
& & 5.	com.ibm.websphere.samples.pbw.jpa.Supplier \\\cline{2-3}

& \multirow{5}{*}{DGI}  & 1. com.ibm.websphere.samples.pbw.ejb.EMailMessage \\\cline{3-3}
& & 2.	com.ibm.websphere.samples.pbw.war.LoginInfo \\\cline{3-3}
& & 3.	com.ibm.websphere.samples.pbw.jpa.Supplier \\\cline{3-3}
& & 4.	com.ibm.websphere.samples.pbw.utils.ListProperties \\\cline{3-3}
& & 5.	com.ibm.websphere.samples.pbw.war.AccountBean \\
\hline

\multirow{15}{*}{Acme-Air} & \multirow{5}{*}{CO-GCN}  &1. com.acmeair.service.ServiceLocator \\\cline{3-3}
& & 2.	com.acmeair.mongo.services.BookingServiceImpl \\\cline{3-3}
& & 3.	com.acmeair.AirportCodeMapping \\\cline{3-3}
& & 4.	com.acmeair.loader.BookingLoader \\\cline{3-3}
& & 5.	com.acmeair.mongo.services.CustomerServiceImpl \\\cline{2-3}

& \multirow{5}{*}{GCN}  &  1. com.acmeair.loader.Loader \\\cline{3-3}
& & 2.	com.acmeair.config.AcmeAirConfiguration.ServiceData \\\cline{3-3}
& & 3.	com.acmeair.web.BookingsREST \\\cline{3-3}
& & 4.	com.acmeair.config.LoaderREST \\\cline{3-3}
& & 5.	com.acmeair.mongo.services.BookingServiceImpl \\\cline{2-3}

& \multirow{5}{*}{DGI}  & 1. com.acmeair.config.AcmeAirConfiguration.ServiceData \\\cline{3-3}
& & 2.	com.acmeair.mongo.services.AuthServiceImpl \\\cline{3-3}
& & 3.	com.acmeair.service.AuthService \\\cline{3-3}
& & 4.	com.acmeair.service.ServiceLocator \\\cline{3-3}
& & 5.	com.acmeair.service.FlightService \\
\hline

\multirow{15}{*}{DietApp} & \multirow{5}{*}{CO-GCN}  &1. WebApplication1.Controllers.ManageController \\\cline{3-3}
& & 2.	WebApplication1.Controllers.ProductController \\\cline{3-3}
& & 3.	WebApplication1.Services.ProductService \\\cline{3-3}
& & 4.	WebApplication1.Services.DietService \\\cline{3-3}
& & 5.	WebApplication1.Models.DietDbContext \\\cline{2-3}

& \multirow{5}{*}{GCN}  &  1. WebApplication1.RouteConfig \\\cline{3-3}
& & 2.	WebApplication1.FilterConfig \\\cline{3-3}
& & 3.	WebApplication1.Services.ProductService \\\cline{3-3}
& & 4.	WebApplication1.BundleConfig \\\cline{3-3}
& & 5.	WebApplication1.Repositories.Concrete.UserRepostiory \\\cline{2-3}

& \multirow{5}{*}{DGI}  & 1. WebApplication1.Controllers.ManageController \\\cline{3-3}
& & 2.	WebApplication1.Services.DietService \\\cline{3-3}
& & 3.	WebApplication1.Repositories.Concrete.UserRepostiory \\\cline{3-3}
& & 4.	WebApplication1.Services.ProductService \\\cline{3-3}
& & 5.	WebApplication1.Controllers.EntryController \\
\hline
\hline
\end{tabular}
\caption{Top five outliers detected using each of the approaches for the four monolith applications}
\label{tab:outliers}
\end{table*}

\begin{table*}[t]
\centering
\begin{tabular}{|c|c|c|c|}\hline
\textbf{Dataset} & \textbf{Annotator} & \textbf{Approaches} & \textbf{\# Overlapping}\\
& & & \textbf{Outlier classes} \\
\hline
\multirow{6}{*}{Daytrader} &  \multirow{3}{*}{Annotator1} & {CO-GCN}  & 3 \\\cline{3-4}
& & {GCN}  & 2 \\\cline{3-4}	
& & {DGI}  & 3 \\\cline{2-4}
& \multirow{3}{*}{Annotator2} & {CO-GCN}  & 3 \\\cline{3-4}
& & {GCN}  & 3 \\\cline{3-4}	
& & {DGI}  & 2 \\\cline{3-4}
\hline
\end{tabular}
\caption{Overlapping outliers for the three approaches from the two annotators for the Daytrader monolith application}
\label{tab:daytrader_outliers_study}
\end{table*}

\begin{table*}[t]
\centering
\begin{tabular}{|c|c|c|c|}\hline
\textbf{Dataset} & \textbf{Annotator} & \textbf{Approaches} & \textbf{\# Overlapping}\\
& & & \textbf{Outlier classes} \\
\hline
\multirow{6}{*}{{PBW}} &  \multirow{3}{*}{Annotator1} & {CO-GCN}  & 3 \\\cline{3-4}
& & {GCN}  & 2 \\\cline{3-4}	
& & {DGI}  & 2 \\\cline{2-4}
& \multirow{3}{*}{Annotator2} & {CO-GCN}  & 3 \\\cline{3-4}
& & {GCN}  & 3 \\\cline{3-4}	
& & {DGI}  & 3 \\\cline{3-4}
\hline
\end{tabular}
\caption{Overlapping outliers for the three approaches from the two annotators  for the PlantsByWebsphere monolith application}
\label{tab:pbw_outliers_study}
\end{table*}

\begin{table*}[t]
\centering
\begin{tabular}{|c|c|c|c|}\hline
\textbf{Dataset} & \textbf{Annotator} & \textbf{Approaches} & \textbf{\# Overlapping}\\
& & & \textbf{Outlier classes} \\
\hline
\multirow{6}{*}{Acme-Air} &  \multirow{3}{*}{Annotator1} & {CO-GCN}  & 2 \\\cline{3-4}
& & {GCN}  & 2 \\\cline{3-4}	
& & {DGI}  & 3 \\\cline{2-4}
& \multirow{3}{*}{Annotator2} & {CO-GCN}  & 3 \\\cline{3-4}
& & {GCN}  & 2 \\\cline{3-4}	
& & {DGI}  & 2 \\\cline{3-4}
\hline
\end{tabular}
\caption{Overlapping outliers for the three approaches from the two annotators for the Acme-Air monolith application}
\label{tab:acmeair_outliers_study}
\end{table*}

\begin{table*}[t]
\centering
\begin{tabular}{|c|c|c|c|}\hline
\textbf{Dataset} & \textbf{Annotator} & \textbf{Approaches} & \textbf{\# Overlapping}\\
& & & \textbf{Outlier classes} \\
\hline
\multirow{6}{*}{DietApp} &  \multirow{3}{*}{Annotator1} & {CO-GCN}  & 2 \\\cline{3-4}
& & {GCN}  & 2 \\\cline{3-4}	
& & {DGI}  & 2 \\\cline{2-4}
& \multirow{3}{*}{Annotator2} & {CO-GCN}  & 4 \\\cline{3-4}
& & {GCN}  & 2 \\\cline{3-4}	
& & {DGI}  & 4 \\\cline{3-4}
\hline
\end{tabular}
\caption{Overlapping outliers for the three approaches from the two annotators for the DietApp monolith application}
\label{tab:dietapp_outliers_study}
\end{table*}

\end{document}